\documentclass{nic-series}

\begin{document} 

\newcommand{\dir}{Figs}

\newcommand{\RR}{{\bf R}}
\newcommand{\rr}{{\bf r}}
\newcommand{\cP}{{\cal P}}
\newcommand{\cN}{{\cal N}}
\newcommand{\cZ}{{\cal Z}}
\newcommand{\cD}{{\cal D}}
\newcommand{\cQ}{{\cal Q}}
\newcommand{\hphi}{\hat{\Phi}}
\newcommand{\bphi}{\bar{\Phi}}

\title{Amphiphiles at Interfaces: Simulation of Structure and Phase Behavior}

\author{Friederike Schmid\inst{1} 
        \and Dominik D\"uchs\inst{1} 
        \and Olaf Lenz\inst{1} 
        \and Claire Loison\inst{1,2}}

\institute{Fakult\"at f\"ur Physik,
        Universit\"at Bielefeld,
        Postfach 100131,
        33501 Bielefeld,
        Germany
        \and
        Centre Europ\'een de Calcul 
        Atomique et Mol\'eculaire, 
        ENS Lyon, 46, All\'ee d'Italie, 
        69007 Lyon, 
        France
        }

\maketitle

\begin{abstracts}
Computer simulations of coarse-grained molecular models for 
amphiphilic systems can provide insight into the
the structure of amphiphiles at interfaces. They can help 
to identify the factors that determine the phase behavior,
and they can bridge between atomic descriptions and 
phenomenological field theories.

Here we focus on model systems for amphiphilic membranes. 
After a brief general introduction, we present selected
simulation results on monolayers, bilayers, and bilayer 
stacks. First, we discuss internal phase transitions in 
membranes and show that idealized models reproduce 
the generic phase behavior. Then we consider membrane 
fluctuations and membrane defects. The simulation data
is compared with mesoscopic theories, and effective
phenomenological parameters can be extracted.
\end{abstracts}

\section{Introduction}

\begin{figure}[t]
\begin{center}
\epsfxsize=3cm
\epsfbox{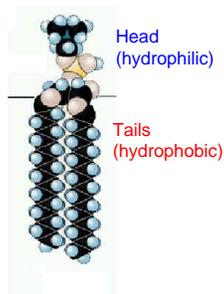}
\caption{\label{fig:amph_molecules}
Structure of a single amphiphilic molecule (DPPC).
}
\end{center}
\end{figure}

\begin{figure}[b]
\begin{center}
\epsfxsize=7cm
\epsfbox{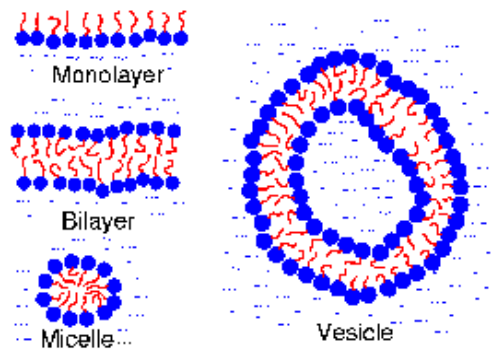}
\caption{\label{fig:amph_structures}
Self-assembled structures of amphiphiles in water.
}
\end{center}
\end{figure}

\begin{figure}[t]
\begin{center}
\epsfxsize=9cm
\epsfbox{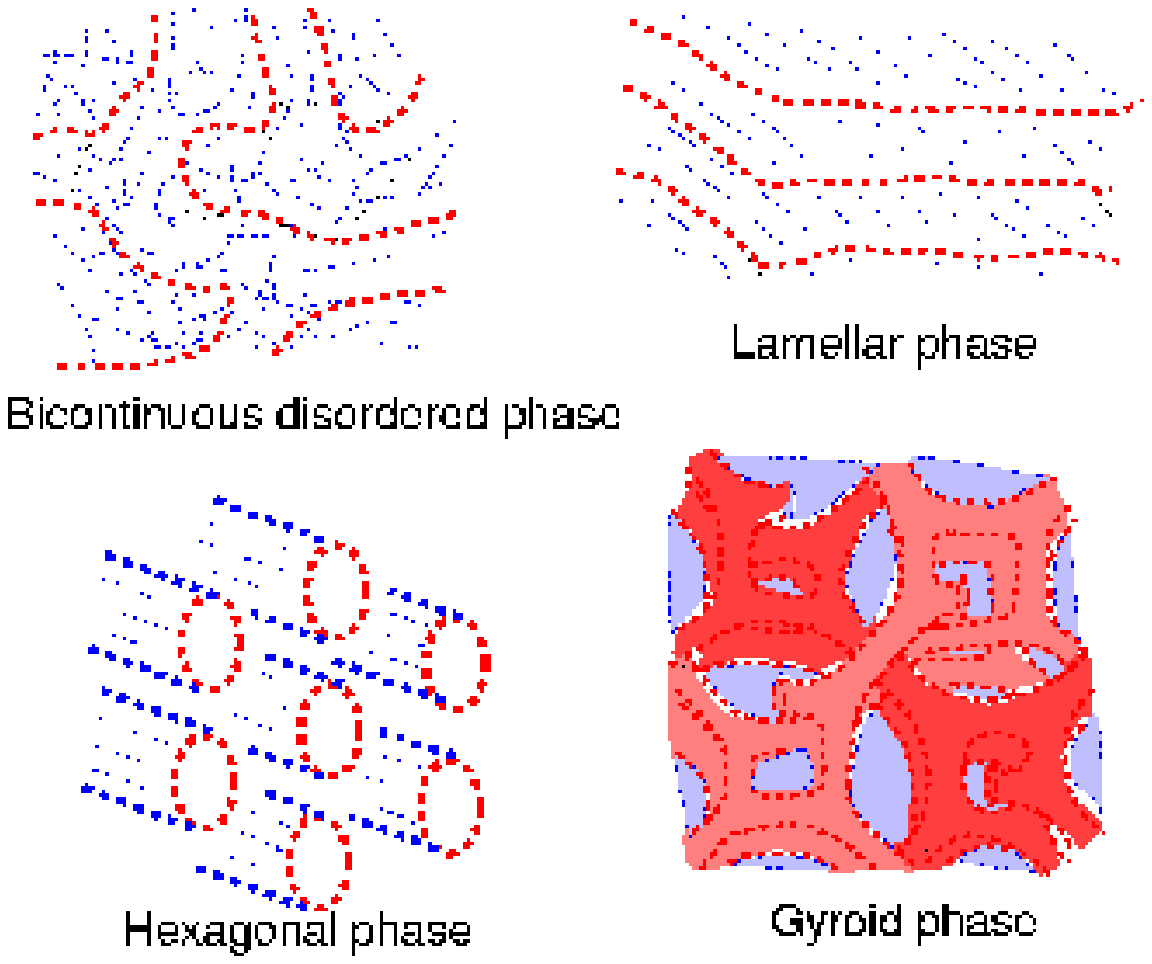}
\caption{\label{fig:amph_phases}
Selected structured phases in amphiphilic systems. 
The dashed lines represent bilayers. The different
shadings in (d) distinguish between two different
interwoven networks in the gyroid phase.
}
\end{center}
\end{figure}

Amphiphilic molecules are characterized by the feature
that they contain both water loving (hydrophilic) 
and water ``hating'' (hydrophobic) structural units. 
Familiar examples are alcohols
or lipids (see Fig.~\ref{fig:amph_molecules}).
Amphiphiles are important for many applications in 
technology and nature. They are very effective at 
helping to dissolve different substances in water, which 
makes them very useful, e. g., as detergents, or as coating
materials to stabilize colloidal systems. Furthermore, they form a 
rich variety of structures at higher concentrations. 
In order to shield their hydrophobic parts from the water,
the amphiphiles self-assemble into spherical or cylindrical 
micelles or bilayers~\cite{israelachvili} 
(Fig.~\ref{fig:amph_structures}). 
These structures may then order on an even higher level and 
form superstructures 
(Fig.~\ref{fig:amph_phases})~\cite{gerhard_review,schubert}. 

Particular interesting from an application point of view are 
those phases where the material is filled with bilayers.
Such bilayer interfaces can serve as barriers against the diffusion
of particles, and help to divide the space into compartments.
Indeed, lipid bilayers are the structural basis of all biological 
membranes, which in turn play a central role for the function of 
all cells and cell organelles~\cite{gennis}.

\begin{figure}[b]
\begin{center}
\epsfxsize=6cm
\epsfbox{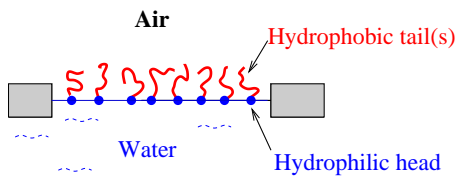}
\caption{\label{fig:langmuir}
Schematic picture of a langmuir trough.
}
\end{center}
\end{figure}

From an experimental point of view, studying the properties of 
biomembranes on a molecular scale {\em in situ} is not an easy task.
Therefore, several model membrane systems have been developed: 
(i) monolayers at the air-water interface (Langmuir monolayers),
(ii) stacks of bilayers, (iii) single planar bilayers, and 
(iii) giant vesicles.
The simplest and oldest approach is to spread lipid molecules 
on a water surface~\cite{kaganer_review}. Due to the amphiphilic 
nature of lipids, they form oriented monolayers at air/water 
interfaces. These are traditionnally placed in a Langmuir trough 
with a movable barrier on one side, which allows to control the 
surface area (see Fig.~\ref{fig:langmuir}). 
The monolayers are then studied in the microscope, with $X$-ray 
scattering under conditions of grazing incidence, or simply by 
measuring the pressure-area isotherms. Since bilayers consist of 
two weakly coupled monolayers, many important bilayer properties 
can be studied already in monolayer systems.

Another experimental approach to studying lipid bilayer membranes 
has been to examine lamellar stacks of bilayers, i. e., lamellar 
phases of lipids~\cite{antelmi,munster,salditt}. Highly aligned 
stacks have been prepared experimentally and then examined by $X$-ray 
diffraction~\cite{munster,salditt}. Such systems have also been useful 
to study the interactions of lipid membranes with other molecules,
e.~g., lipid-polymer interactions~\cite{antelmi2}, 
lipid-protein interactions~\cite{koltover,munster2}, 
lipid-DNA interactions~\cite{radler1,radler2}.

Furthermore, there also exist techniques by which planar bilayers or 
giant bilayer vesicles can be generated. $X$-ray studies of such
systems are difficult, due to the small size of the sample.
However, they can be studied by other techniques, such as phase 
contrast microscopy or electron microscopy, light scattering, 
or transport measurements~\cite{gennis,doebereiner}.
It should be noted that isolated membranes are not stable in a 
thermodynamical sense. They are metastable structures,
bound to break eventually. This, however, does not
restrict their importance as model systems for biological
membranes, which are themselves nonequilibrium structures.

When attempting to describe lipid membranes theoretically, one 
faces a problem that is common for soft materials: The membrane 
properties result from an interplay between physical and 
chemical phenomena that live on a hierarchy of length scales, 
ranging from the atomic to the mesoscopic scale (micrometers), 
and likewise on a hierarchy of time scales. 
On the scale of atoms and small molecules (water, ions), 
one has the forces which keep the membrane together 
in the first place: the hydrophobic effect and the interaction 
between water and hydrophilic head groups, which is still not 
yet fully understood. 
On the molecular scale, one has the interplay between local
chain conformations, membrane structure, and membrane viscosity 
and fluidity. We will come back to that aspect further below. 
One also has the electrostatic interactions between lipids. 
For bilayers under physiological conditions, they are shielded to 
some extent by the ions in the surrounding water. 
Nevertheless, they still influence the effective size of the head
groups, thus stabilizing or destabilizing the planar lamellar 
structure. In monolayers, electrostatic interactions are
not shielded and generate mesoscopic patterns~\cite{connell,knobler}.
On a supramolecular length scale, one has the fluctuations
of membrane positions. These are often described in terms 
of phenomenological parameters such as the bending stiffness and 
(if applicable) the surface tension~\cite{gerhard_review,safran}. 
Finally, on the mesoscopic length scale, one has phase separation
and domain formation. This becomes particularly important in 
{\em mixed} systems.  For example, phase separation of different 
lipids within a membrane may trigger budding 
processes~\cite{lipowsky,doebereiner2,yamamoto}.
Certain membrane lipids and proteins aggregate into ``rafts'', 
which are believed to serve a biological purpose~\cite{brown}. 
Polymers may induce membrane fusion~\cite{safran2}.

Computer simulations which cover all these aspects are not possible, 
and will not be possible for a long time. When studying membranes 
by computer simulation, the first and main task is therefore to
{\em choose a problem} and then {\em identify an appropriate model}.
Different models operate on different length and time scales. 
Ideally, it should be possible to establish connections between 
the different models (``multiscale modeling''). The problem of 
bridging between time and length scales is one of the great 
challenges in today's computational science. We are still far 
from having reached that goal. However, there exist several models 
which can be used to study membranes on particular length scales, 
and computer simulations of these models can contribute to an 
improved understanding of physical phenomena on that scale. 

In fact, several approaches are introduced in this school. 
Ole Mouritsen presents lattice models which allow to study 
the mesoscopic organization of complex biomembranes. 
Alan Mark takes the opposite perspective and studies membranes 
on an atomistic level. Here, we will assume an intermediate point
of view and discuss molecular coarse grained models.  
Such models can be used to study the structure of a system
systematically for a range of parameters. For example, they
reproduce self-assembly and different lipid-water phases, as well 
as internal phase transitions in lipid membranes. Numerical 
studies provide insight into mechanism which contribute to the 
phase behavior.  Moreover, molecular coarse grained models can 
be used as starting points for simulations that bridge between 
the molecular and the mesoscopic level. An example of such a 
study will be presented in Sec.~\ref{sec:stacks}.

The paper is organized as follows. We will begin with a brief
discussion of typical models that are used to study membranes.
In Sec.~\ref{sec:applications}, we present some applications of 
coarse grained molecular models to study monolayers, bilayers, 
and bilayer stacks. We conclude in Sec.~\ref{sec:outlook}.

\section{Models and Levels of Coarse Graining}
\label{sec:model}

As discussed in the introduction, it is not possible to define 
a single simulation model that is suited to study all relevant 
aspects about amphiphile membranes in one single computer simulation. 
Therefore, a hierarchy of models has been developed and used to 
study membranes from different point of views.

The models at the lowest level of the hierarchy are the atomistic 
models. Even these are not truly {\em ab-initio} in the
sense that the molecules are not treated in full quantum chemical 
detail. Atomistic molecular dynamics simulations of amphiphilic 
systems use semiempirical potentials such as the GROMOS force 
field (see Alan Mark's contribution), which are constructed by 
fitting the parameters to quantum chemical calculations {\em and} 
to experimental data. Thus they focus on the realistic description 
of specific substances. The time and length scales accessible to 
such a simulation are limited, but they increase very rapidly due 
to the improving performance of the computer hardware, and to 
the development of good algorithms. Nowadays, atomistic simulations 
can deal with thousands of amphiphiles, or reach time scales up to 
microseconds.

The next level of coarse graining is that of idealized molecular 
models. Here, different molecules are still treated as individual 
particles, but their structure is very much simplified. Atomic and 
molecular details are largely disregarded. Only the features that 
are essential for the behavior of the molecules are kept. In the 
case of amphiphiles, one characteristic attribute is clearly the 
dual character of the molecules, with their separate water-loving 
and water-hating parts. Another quantity that seems to be important 
for the self-organization of the amphiphiles is the conformational 
entropy of the molecules.  Many coarse-grained amphiphile models 
concentrate precisely on these two aspects of the molecules, and 
represent amphiphiles by chains of simple ``water''-loving or 
``water''-repelling units. 
Coarse grained models are particularly suited to study {\em generic} 
properties of amphiphiles. They can be regarded as simple, minimal 
systems that provide general insight into the mechanisms that drive 
the self-assembly and the phase behavior of amphiphiles. 

To optimize computational efficiency, many coarse grained models 
have been formulated on a lattice. A particular popular lattice 
model has been introduced about twenty years ago by R.~G. Larson 
{\em et al.}~\cite{larson}. Water molecules ($w$) occupy 
single sites of a cubic lattice, and amphiphile molecules are modeled 
by chains of ``tail'' monomers $t$ and ``head'' monomers $h$, which 
are taken to be identical to $w$-particles. Only particles that are 
neighbors on the lattice interact with each other. The lattice is 
entirely filled by $w$, $h$, or $t$ particles. The interaction 
energy is thus determined by a single interaction parameter, 
which describes the relative repulsion between $t$ and $w$ or $h$. 
The model can be simulated very efficiently by Monte Carlo methods. 
It exhibits an amazingly rich phase behavior~\cite{ted}. 
Even a gyroid structure can be observed under certain 
circumstances~\cite{larson_gyroid}.

Despite the success of lattice models, they still have the obvious 
flaw of imposing an {\em a priori} anisotropy on space. Therefore, 
off-lattice models are attracting growing interest. To our best 
knowledge, the first amphiphile model of this kind was introduced 
by B.~Smit {\em et al.}~\cite{smit1} in 1990. The general idea is 
similar to that of R.~G. Larson. The amphiphiles are 
represented by chains made of very simple $h$- or $t$-units, which 
are in this case spherical beads. The ``water'' molecules are 
represented by free beads. Beads interact via simple short-range 
pairwise potentials, often truncated Lennard-Jones potentials. 
The parameters of the potentials are chosen such that $ht$ pairs 
and $hw$ pairs effectively repel each other. Such a model reproduces
self-assembly, micelle and membrane formation.

Many similar amphiphile models have been defined 
and applied to study various problems. Some examples from 
our own work will be presented in the next section
(Sec.~\ref{sec:applications}).

The models discussed so far still treat amphiphiles
as flexible chains. One might argue that the chain
flexibility is not absolutely essential for the
character of an amphiphilic system, and that a truly 
``minimal'' model should ignore it. Indeed, ``molecular''
models that restrict themselves to the very basic ingredients
have been efficient tools to study certain aspects
of amphiphilic self-organization. For example, spin
models which include just the orientation of amphiphiles 
and the repulsion between one molecule end and water have 
reproduced topological characteristics of amphiphilic
phase behavior~\cite{gerhard_review}. A particularly successful 
class of lattice models has been designed specifically 
to model lipid membranes~\cite{pink1,pink2,bernd}, and has been 
used to study various complex biomembranes at equilibrium 
and even nonequilibrium. This approach will be presented 
in Ole Mouritsen's lecture.

Finally, the highest level of coarse graining is reached
with the phenomenological models. These drop the notion of 
single particles entirely and describe amphiphilic systems 
by continuous fields, with a free energy functional that
is governed by a few mesoscopic material parameters. 
For example, Ginzburg-Landau models~\cite{gerhard_review}
introduce a free energy functional, which depends
on local amphiphile and water concentrations. 
In contrast, random interface models~\cite{safran,gerhard2}
concentrate on the amphiphilic sheets, which are 
parametrized and characterized in terms of mechanical
elasticity parameters such as the bending moduli.
Mesoscopic models are good starting points for analytical
approaches. Thus computer simulations of such models may 
also serve to test or to complement an analytical theory.

After this brief overview over different types of models 
for amphiphilic systems, we will now focus on coarse 
grained molecular models. In the next section, we will 
show how computer simulations of such models can help 
to understand the structure and the phase behavior of 
amphiphilic monolayers and bilayers.

\section{Applications to Amphiphilic Layers}
\label{sec:applications}

This chapter will address two different aspects of the 
structure of amphiphilic layers: Internal phase transitions, 
and fluctuations and defects in membrane stacks. 
It is by no means a complete overview over all simulation
studies that have dealt with these and related issues.
Instead, it mostly focusses on some of our own work, which 
is hopefully suited to illustrate the potential and the
merits of molecular coarse-grained simulations.

Lamellar stacks of lipids in water assume several different
structures. At high temperatures and low pressures, they are usually 
in the ``fluid'' $L_{\alpha}$ phase, which is characterized
by a low degree of conformational order and a high mobility
of lipids within the membranes. Upon lowering the temperature, 
one encounters a first order transition -- the ``main transition'' --
to a state with higher conformational order, and lower mobility: 
a ``gel'' state. In the gel region, there exist different 
phases. For example, the chains may display collective tilt with
respect to the surface normal, they may show local hexatic
order, the lamellae may even exhibit asymmetric wavy 
undulations (ripple phase). Most biomembranes in living 
organisms are maintained in the fluid state. Nevertheless, 
the main transition has presumably some relevance for 
biological systems, as it occurs at temperatures close to the 
body temperature for some common bilayer lipids 
({\em e. g.}, 41 ${}^0$ in DPPC). It will be considered
in the subsections \ref{sec:monolayers} and
\ref{sec:bilayers}.

The internal ordering of the lipids is one important structural
property of a membrane. Another is the spectrum of shape fluctuations, 
and the frequency and structure of membrane defects. 
Membrane defects determine critically the permeability of 
membranes. Moreover, the formation of point defects is believed 
to play a crucial role as a first step in the process of membrane 
fusion~\cite{marcus1}.
Membrane fluctuations and membrane defects have been described with 
phenomenological approaches~\cite{holyst,lei,lister,shillcock},
and these theories were used to interpret experimental results.
In computer simulations, the local structure can be investigated 
in much more detail than in experiments. Therefore comparisons
of phenomenological theories with molecular simulations are
clearly of interest. An example of such a comparison will be presented
in the subsection \ref{sec:stacks}.

\subsection{Monolayers}
\label{sec:monolayers}

We begin with discussing the phase behavior in monolayers.
It is closely related to that of bilayers, which is one of the 
reasons why Langmuir monolayers are considered to be useful 
model membrane systems. In particular, the main transition has a 
prominent monolayer equivalent -- a first-order phase transition 
between a ``liquid expanded'' state and a denser ``liquid condensed'' 
state. As in membranes, the condensed state exists in several 
modifications, which differ, among other, by tilt order and 
positional order of the head groups, or by backbone order of 
the chains.

\begin{figure}[b]
\begin{center}
\epsfxsize=6cm
\epsfbox{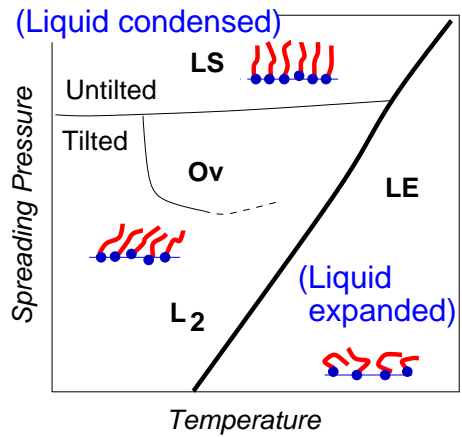}
\caption{\label{fig:fatty_acids}
Generic phase diagram for fatty acid monolayers.
(after Ref.~\citeonline{kaganer_review}). LE is the
liquid expanded phase. In the L${}_2$ phase and the
Ov phase, the chains are tilted towards next-nearest
and nearest neighbors, respectively. In the LS phase,
they are on average untilted. Further phases are
found at lower temperatures (not shown).
}
\end{center}
\end{figure}

Experimentally, monolayer phase diagrams are often determined
as a function of the temperature and the area per molecule,
or alternatively, the spreading pressure of the molecules
on the surface. The phase diagrams for different amphiphilic
molecules are very similar. As an example, 
Fig.~\ref{fig:fatty_acids} shows a generic phase diagram
of fatty acid monolayers~\cite{kaganer_review,bibo,peterson,overbeck}. 

In order to describe such systems on a coarse-grained level,
we model the amphiphiles as chains of $N$ ``tail'' beads with 
diameter $\sigma_t$, attached to a slightly larger ``head'' 
bead of diameter $\sigma_h > \sigma_t$. The water surface
is represented by a planar surface at $z=0$, which repels the
tail beads, and attracts the head bead. 

More specifically, the model is defined as follows:
The beads in the chains are connected by nonlinear springs
with the potential
\begin{equation}
\label{fene}
V_{S}(d) = \left\{ \begin{array}{l c r}
- \frac{k_{S}}{2} \; d_{S}^2 \; \ln\Big[ 1 - (d-d_0)^2/d_{S}{}^2 \Big]
\quad & \mbox{for} & |d-d_0|<d_{S} \\
\infty & \mbox{for} & |d-d_0| > d_{S}
\end{array} \right.,
\end{equation}
where $d$ is the length of the spring.
This potential is constructed such that it is nearly harmonic
in the center, at $d \approx d_0$, but diverges at $d = d_S$.
Thus the distance between neighbor beads cannot take arbitrarily
large values, which ensures that chains cannot cross each other.
A potential with a logarithmic cutoff as in (\ref{fene}) is often 
called FENE-potential (Finite Extendible Nonlinear Elastic potential).
In addition to this spring potential, chains are given a
bending stiffness by virtue of a stiffness potential
\begin{equation}
\label{ba}
V_{A} = k_{A} \cdot (1- \cos \theta),
\end{equation}
which acts on the angle $\theta$ between subsequent springs and favors
$\theta = 0$ (straight chains). The parameter $k_A$ can then be tuned to 
control the conformational freedom of the chains and study the role of 
the chain entropy for the phase behavior.

Two beads $i$ and $j$ ($i,j = h$ or $t$) that are not direct neighbors
in the same chain interact with a truncated and shifted Lennard-Jones 
potential,
\begin{equation}
\label{vlj}
V_{ij}(r) = \left\{ \begin{array}{lcr}
\epsilon \:
\Big( \big[\sigma_{ij}/r\big)^{12} - 2 \big(\sigma_{ij}/r\big)^6 + v_{ij} \Big] 
\quad & \mbox{for} & r \le R_{ij} \\
0 & \mbox{for} & r > R_{ij}
\end{array} \right. ,
\end{equation}
where $\sigma_{ij} = (\sigma_i + \sigma_j)/2$, the cutoff is
$R_{tt} = 2 \sigma_{tt}$ for the tail beads, and $R_{ht}=\sigma_{ht}$,
$R_{hh}=\sigma_{hh}$ for the other interactions, and the shifting
parameter $v_{ij}$ is chosen such that $V_{ij}(r)$ is continuous
at $r = R_{ij}$. It is easy to see that with this choice of the cutoff 
parameters, the interactions between tail beads are attractive,
and all other interactions are purely repulsive.

Finally, we have to specify the interactions of beads with the
surface.  We have explored two types of surface potentials: 
First we used a model introduced by Haas {\em et al}~\cite{haas}, 
where the surface imposes rigid constraints: The head beads are
confined to stay in the plane at $z=0$, and the tail beads are
not allowed into the half-space $z<0$~\cite{christoph1,christoph2}. 
This model has the advantage of being very simple, however, it
is unphysical in the sense that real water surfaces are not sharp 
on an atomic scale. Indeed, better agreement with the experimental 
phase diagram can be reached with softer surface 
potentials~\cite{dominik}. Here, head beads are subject to
the potential
\begin{equation}
\label{vh}
V_{h}(r) = \left\{ \begin{array}{lcr}
0 & \mbox{for} & z < -0.5 W\\
-{\epsilon_s}/{2}\: \ln \Big[ 1-(z+0.5 W)^2/W^2 \Big]
\quad & \mbox{for} & -0.5 W < z < 0.5 W 
\end{array} \right. ,
\end{equation}
and tail beads to
\begin{equation}
\label{vt}
V_{t}(r) = \left\{ \begin{array}{lcr}
-{\epsilon_s}/{2}\: \ln \Big[ 1-(z-0.5 W)^2/W^2 \Big]
\quad
& \mbox{for} & -0.5 W < z < 0.5 W \\
0 & \mbox{for} & z > 0.5 W\\
\end{array} \right. .
\end{equation}
with $W = 1 \sigma_t$ and $\epsilon_s = 10 \epsilon$. It turns out
that the exact form of the surface potential is not important,
almost identical results are obtained with $W = 2 \sigma_t$.

The other model parameters were $d_0 = 0.7 \sigma_t$, $d_S= 0.2 \sigma_t$, 
and $k_S = 100 \epsilon$. The stiffness parameter $k_A$ was varied between 
$4.5 \epsilon$ and $100 \epsilon$. The head size was chosen 
$\sigma_h = 1.1 \sigma_t$ or $\sigma_h = 1.2 \sigma_t$, and the chain 
length was $N=7$ in most simulations.

\begin{figure}[t]
\begin{center}
\epsfxsize=10cm
\epsfbox{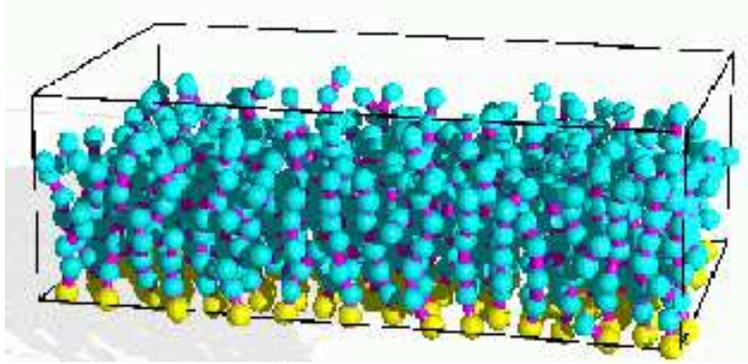}
\caption{\label{fig:ml_snapshot}
Snapshots of a monolayer in a system with $\sigma_h = 1.1 \sigma_t$,
$k_A = 10 \epsilon$ and rigid constraints at the surface.
($T=4 \epsilon/k_B$ and $P = 50 \epsilon/\sigma_t^2$) 
(Courtesy of C. Stadler).
}
\end{center}
\end{figure}

This model was studied by Monte Carlo simulations at constant spreading 
pressure $P$: We considered $n$ chains with heads grafted in a planar 
parallelogram of variable size and shape, characterized by two
side lengths $L_x$ and $L_y$ and one angle $\alpha$. 
The boundary conditions were periodic in the $xy$ plane, and free in 
the $z$ direction. The simulation algorithm included three elementary
Monte Carlo moves:
\begin{itemize}
\item Single particle displacements
\item Variations of $L_x$ or $L_y$, {\em i.~e.}, the whole configuration
   is stretched or squeezed in one direction. Care must be taken that
   this move satisfies detailed balance. For example, a move of the 
   form $L \to L \delta$, where $\delta$ is uniformly distributed,
   cannot be applied without introducing a correction term in the 
   probability for accepting the move. In contrast, the move 
   $L \to L \exp(\delta)$ can be applied without correction. 
   In our simulations, we varied $L$ in an additive way, 
   $L \to L + \delta$ (rejecting moves that made $L$ negative).
\item Shearing the box, {\em i.~e.}, variation of $\alpha$ while
   keeping the total area constant.
\end{itemize}
The moves were accepted or rejected according to the Metropolis
prescription~\cite{binder_heermann,landau_binder,frenkel_smit} 
with the effective Hamiltonian
\begin{equation}
{\cal H} = E + P A - n N T \log A,
\end{equation}
where $E$ is the (internal) energy, and $A$ is the area of the system.
To speed up the simulations, we used cell lists and Verlet 
lists~\cite{allen_tildesley,frenkel_smit}.
Fig.~\ref{fig:ml_snapshot} shows an example of a configuration snapshot.

\begin{figure}[t]
\begin{center}
\epsfxsize=6cm
\epsfbox{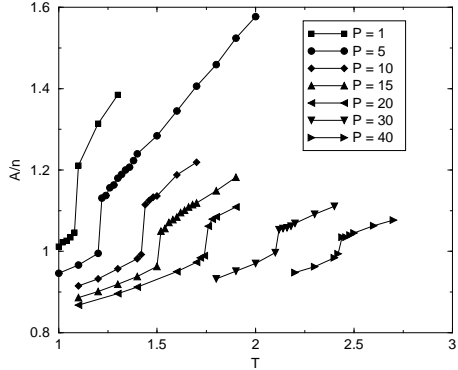}
\caption{\label{fig:at}
Area per molecule (in units $\sigma_t^2$) vs. temperature 
(in units $\epsilon/k_B$) in a system with $k_A = 4.7 \epsilon$,
$\sigma_h = 1.1 \sigma_t$ and soft surface potentials.
The spreading pressures are given in units of $\epsilon/\sigma_t^2$.
From Ref.~\citeonline{dominik_diplom}.
}
\end{center}
\end{figure}

To analyze the properties of the monolayer, one needs to define
and monitor appropriate observables. One quantity of interest is the 
area per molecule. Fig.~\ref{fig:at} shows typical temperature-area 
isobars. One clearly discerns a discontinuous jump, which moves to 
higher temperatures with increasing pressure.

\begin{figure}[t]
\begin{center}
\epsfxsize=6cm
\epsfbox{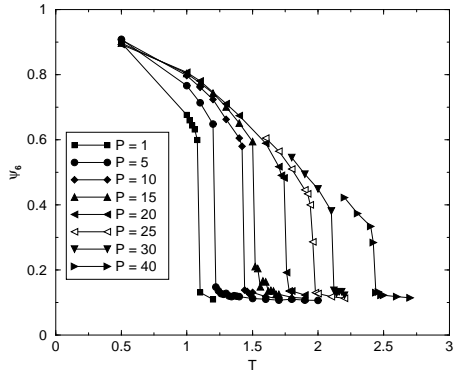}
\caption{\label{fig:psi6}
Order parameter $\psi_6$ vs. temperature (units $\epsilon/k_B$)
for the same system as Fig.~\ref{fig:at}.
From Ref.~\citeonline{dominik_diplom}.
}
\end{center}
\end{figure}

The nature of this phase transition can be studied in more detail
by inspecting the radial pair correlation functions and the
hexagonal order parameter of two-dimensional melting,
\begin{equation}
\label{psi}
\Psi_6 = \bigg\langle \bigg| \frac{1}{6n}
\sum_{j=1}^n \sum_{k=1}^6 \exp(i 6 \phi_{jk}) \bigg|^2 \Big\rangle.
\end{equation}
Here the sum $j$ runs over all heads of the system, the sum $k$ over
the six nearest neighbors of $j$, and $\Phi_{jk}$ is the angle
between the vector connecting the two heads and an arbitrary reference
axis. The data for $\Psi_6$ which correspond to Fig.~\ref{fig:at}
are shown in Fig.~\ref{fig:psi6}. The hexagonal order parameter
is nonzero in the low temperature phase, and jumps to a value
close to zero at the transition. This indicates that the high
temperature phase is liquid and the low temperature phase is
hexatic or (quasi)crystalline.

\begin{figure}[t]
\begin{center}
\epsfxsize=6cm
\epsfbox{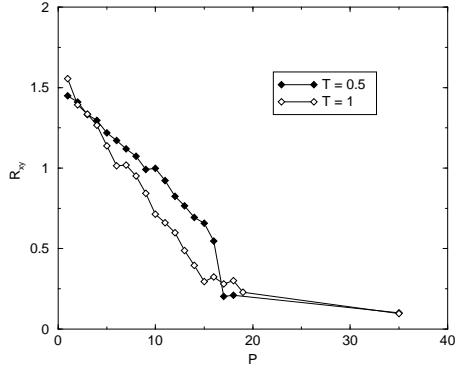}
\caption{\label{fig:rxy}
Tilt parameter $R_{xy}$ vs. pressure (in units $\epsilon/\sigma_t^2$)
at two temperatures $T$ (in units $\epsilon/k_B$)
for the same system as Fig.~\ref{fig:at}.
From Ref.~\citeonline{dominik_diplom}.
}
\end{center}
\end{figure}

A closer inspection of the ordered low-temperature region reveals 
that it can be subdivided in at least two regimes: At low pressures, 
the chains are collectively tilted in one direction, and at high
pressure, they are on average untilted. The azimuthal symmetry breaking 
due to collective tilt order can be characterized by the order parameter
\begin{equation}
R_{xy} = \sqrt{\langle [x]^2 + [y]^2 \rangle },
\end{equation}
where $[x]$ and $[y]$ denote the $x$ and $y$ components, respectively,
of the head-to-end vector of a chain, averaged over all chains in {\em one}
configuration, and $\langle \cdot \rangle$, the statistical average
over all configurations. The values of $R_{xy}$ as a function of 
the pressure on two different isotherms are shown in Fig.~\ref{fig:rxy}.

\begin{figure}[b]
\begin{center}
\epsfxsize=7cm
\epsfbox{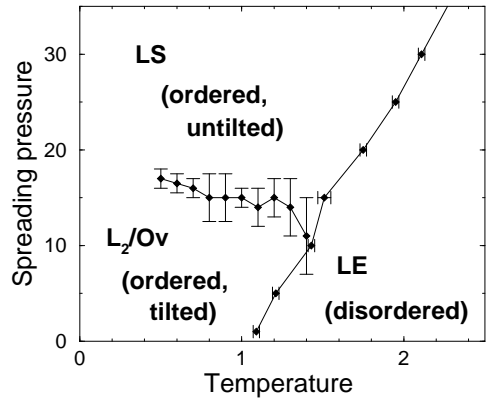}
\caption{\label{fig:ml_phase_diagram}
Phase diagram from Monte Carlo simulations of the monolayer model with
$k_A = 4.7 \epsilon$, $\sigma_h = 1.1 \sigma_t$ and soft surface potentials.
soft surface potentials in the plane of spreading pressure 
(units $\epsilon/k_B \sigma_t^2$) vs. temperature (units $\epsilon/k_B$).
The chain stiffness is $k_A = 4.7 \epsilon$. LE denotes disordered
phase, LS untilted ordered phase, and L${}_2$/Ov ordered phase with
tilt towards nearest or next-nearest neighbors (undetermined).
From Refs.~\citeonline{dominik,dominik_diplom}
}
\end{center}
\end{figure}

The data for different temperatures and spreading pressures can
be summarized in a phase diagram. An example for the system with 
soft potentials is shown in Fig.~\ref{fig:ml_phase_diagram}. 
This phase diagram has great similarity with the high temperature
part of the experimental phase diagram in Fig.~\ref{fig:fatty_acids}.
At high temperature, we observe a disordered liquid expanded phase (LE).
At low temperature two types of condensed phases are present -- 
an untilted high-pressure phase (LS) and a tilted low-pressure
phase (L${}_2$/Ov). The temperature of the order-disorder transition 
increases with pressure, as in the experiment, and the pressure of the
tilting transition between LS and L${}_2$/Ov is almost independent
of the temperature, again as in the experiment.

Thus we have established a minimal model which reproduces the
main features of Langmuir monolayers in the vicinity of the
main transition. At lower temperatures, the experimental phase 
diagram of fatty acids is much more complex than that of our model.
This is not surprising. To reproduce phenomena like backbone
ordering, the chains have to be modeled in much more detail than 
has been done here. On the other hand, the details of the low 
temperature portion of the experimental phase diagram depend 
strongly on the particular choice of the amphiphile.
Our model was never designed to describe specific properties
of fatty acids. It was designed to reproduce the main transition
in amphiphile monolayers. This is achieved in a satisfactory way.

We can learn more about the main transition by playing with 
the model parameters and studying how this influences the phase 
behavior. For example, increasing the chain stiffness shifts the 
order/disorder transition to higher temperatures and the tilting 
transition to higher pressures~\cite{christoph1}. The form of the 
surface potential (rigid or soft) is not important for the 
order/disorder transition, but it does influence the slope of the 
line of tilting transitions in pressure-area space~\cite{dominik}.
Likewise, increasing the head group size by 10 \% does not 
influence the order/disorder transition very much, but may
affect the tilted phases quite dramatically~\cite{christoph2}. 
All these observations can be summarized by the statement
that the order/disorder transition is basically driven by 
the chains, whereas the tilting transitions result from 
an interplay between chains and head groups. 

\subsection{Bilayers}
\label{sec:bilayers}

As mentioned earlier, lipid bilayers exhibit internal phase 
transitions which are very similar to those observed in monolayers. 
In other respect, however, they are fundamentally different. 
They do not form by adsorption on a pre-defined surface, instead, 
the lipids self-assemble spontaneously to a structure with 
planar geometry. As a result, bilayers are not strictly planar, 
but may curve around and undulate. In order to study membranes,
one needs to define a model which reproduces self-assembly.

How can bilayer self-assembly be modeled? The simplest approach 
is to force amphiphiles into sheets by tethering the head groups 
to two dimensional surfaces~\cite{harries,baumg1,baumg2,sintes}. 
``Self-assembly'' is then enforced by external constraints,
with the obvious consequence that the head groups lose much
of their translational degrees of freedom. In contrast, real 
self-assembled structures are held together by an interplay of 
amphiphile-amphiphile and amphiphile-solvent interactions. 
Coarse-grained models that reproduce spontaneous self-assembly
must account for the presence of solvent one way or another. 
However, the solvent should be modeled with as little 
detail as possible, since the focus is still on the bilayers.

One possibility is to use the Smit model~\cite{smit1} mentioned 
in the introduction. The amphiphiles are modeled by chains of 
beads, and the solvent by beads of the same size. This model 
indeed reproduces bilayer self-assembly~\cite{goetz1} and has 
been applied successfully by Goetz {\em et al.} to study shape 
fluctuations of model bilayers~\cite{goetz2}. 

Unfortunately, even the simple Smit model still has the drawback 
that a substantial amount of computer time in simulations is spent 
on the uninteresting solvent. It is therefore desirable to have a 
model which does not include the solvent explicitly. This idea is 
not particularly eccentric, such models are commonly used in 
simulations of polymers in solvent. The effect of the solvent 
is then incorporated in the effective interactions between monomers. 
However, it is not {\em a priori} clear that effective (pair) 
interactions will be able to bring about something as complex as 
membrane self-assembly. 

Indeed, it was only very recently that O. Farago~\cite{farago} 
established a solvent-free molecular model for membranes.
Amphiphiles are represented by rigid linear trimers, made
of beads which interact through truncated Lennard-Jones
interactions with carefully chosen interaction parameters.
O. Farago showed that single bilayers remain (meta)stable in
his model, that they exhibit an order-disorder phase
transition as a function of the molecular area density,
and that they even sustain pore formation. However,
he also mentions that a lengthy ``trial and error''
process of fine tuning the Lennard-Jones parameters
was necessary to achieve this impressive result.

\begin{figure}[b]
\begin{center}
\epsfxsize=5.5cm
\epsfbox{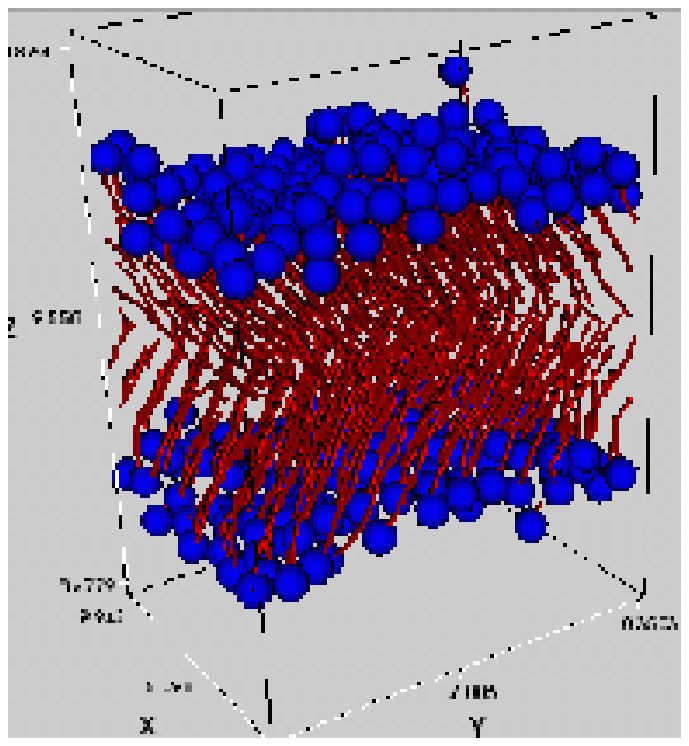}
\hfill
\epsfxsize=5.5cm
\epsfbox{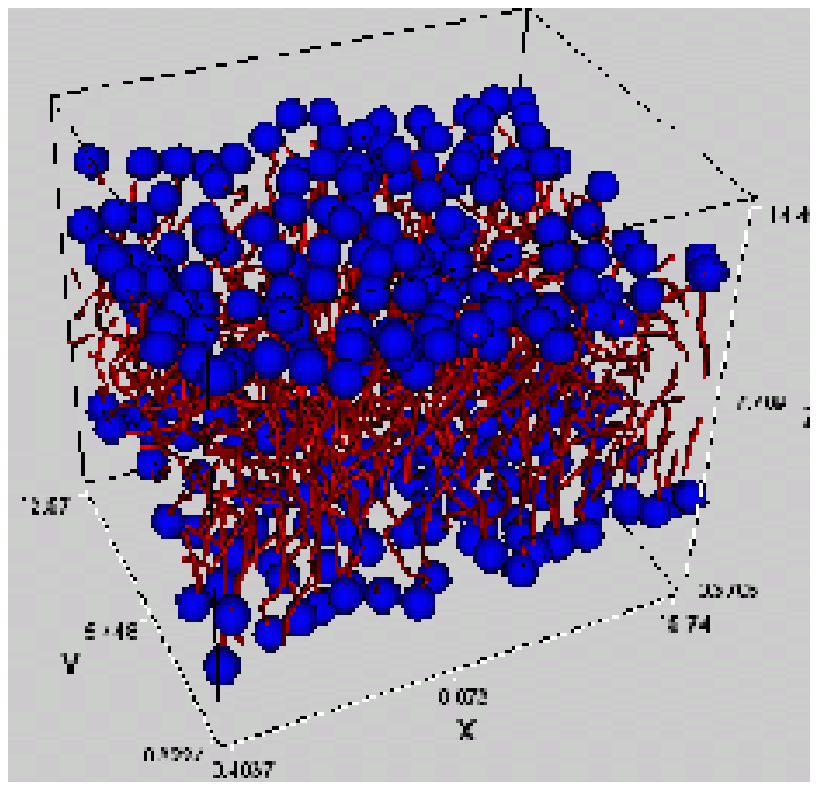}
\caption{\label{fig:bl_snapshot}
Snapshots of bilayers embedded in phantom solvent
at pressure $P = 1 \epsilon/\sigma_t^3$ in the ordered
state (left, temperature $T=0.9 \epsilon/k_B$) and in 
the fluid state (right, $T=1.0 \epsilon/k_B$).
Solvent particles are not shown.
($\sigma_h = 1.1 \sigma_t$, $k_A = 4.7 \epsilon$) 
}
\end{center}
\end{figure}

In our own work~\cite{lenz}, we propose a different, 
more robust approach, the ``phantom solvent'' model.
Solvent particles are treated explicitly, but they 
interact {\em only} with the amphiphiles, not with one 
another. The amphiphiles perceive the solvent particles are 
soft beads of radius $\sigma_s=\sigma_h$. The bulk of the
solvent region is simply an ideal gas. The 
amphiphiles are modeled exactly in the same way as 
in our previous monolayer model, except that the
surface potentials (\ref{vh}) and (\ref{vt}) are of
course eliminated and replaced by periodic boundary
conditions. The model can be implemented in a 
straightforward way and studied by Monte Carlo simulations. 
We found that it produced stable bilayers at the first go, 
without any parameter tuning. Two examples of 
snapshots are shown in Fig.~\ref{fig:bl_snapshot}.

The phantom solvent model has several advantages:
\begin{itemize}
\item It is computationally very efficient. The 
  computational time spent on the phantom solvent
  is only a few percent of that spent on the amphiphiles,
  for any reasonable number of solvent beads. This
  is because only few pair interaction potentials 
  have to be evaluated per solvent bead.
\item The solvent has no local liquid structure. 
  This is good, because we are not interested in the 
  interplay between solvent and bilayer 
  structure. If we intended to study this aspect, 
  we would have to model the solvent (water!)
  in much more detail. Moreover, a structured
  solvent would introduce unwanted effective 
  interactions between the bilayer and it's periodic 
  images in the normal direction.
\item The phantom solvent has a simple physical
  interpretation. It probes the total free volume 
  that is available to the solvent on the length 
  scale $\sigma_s$. The self-assembly in our model 
  is driven by the attractive interactions between
  the tails, and by the entropic effect that the 
  solvents have more space when the amphiphiles 
  are clustered together.
\end{itemize}
At the moment, the model still shares the handicap
of all solvent-free models (and lattice models), that 
hydrodynamical interactions between the membranes and 
the solvent cannot be treated very easily. It might be 
possible to formulate dynamical equations for the phantom 
solvent which ensure that it behaves like a liquid and 
not like a gas. Otherwise, those who intend to study 
hydrodynamic effects might have to equip the solvent 
particles with a weak integrable potential, as is done 
in dissipative particle dynamics simulations.
As long as we calculate static properties with Monte 
Carlo simulations, this is however not a problem.

\begin{figure}[t]
\begin{center}
\epsfxsize=7cm
\epsfbox{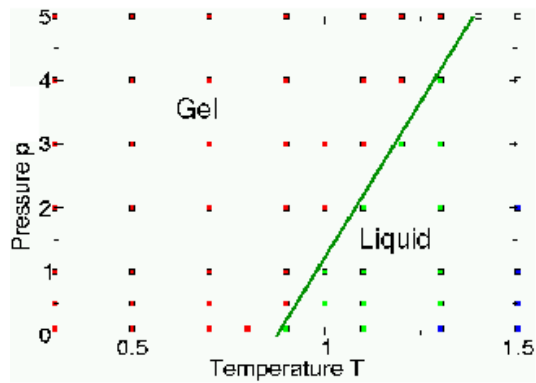}
\caption{\label{fig:bl_phase_diagram}
Preliminary phase diagram of bilayers in the phantom 
solvent model in the plane pressure 
(units $\epsilon/\sigma_t^3$)
vs. temperature (units $\epsilon/k_B$).
The model parameters are the same as in 
Fig.~\ref{fig:ml_phase_diagram}. Red points denote
ordered membrane, and green points fluid membrane.
At high temperatures (blue points), the membrane 
disintegrates. 
}
\end{center}
\end{figure}

After these general remarks, we turn back to the 
discussion of our particular amphiphile model. 
As mentioned before, it produced stable bilayers over
a wide parameter range. 
The configuration snapshots of Fig.~\ref{fig:bl_snapshot} 
demonstrate that the bilayers exhibit a low-temperature 
ordered phase and a high-temperature disordered phase. 
At even higher temperatures, they disintegrate. 
The phase transitions can be monitored, {\em e. g.},
by inspecting the total nematic order parameter of
amphiphiles or the area per lipid as a function of
the pressure and temperature. A preliminary phase
diagram is shown in Fig.~\ref{fig:bl_phase_diagram}.
We note the similarity to the monolayer phase diagram
of Fig.~\ref{fig:ml_phase_diagram}. This corroborates 
the assertion that monolayers are good model systems 
for membranes.

\subsection{Bilayer Stacks}
\label{sec:stacks}

Having discussed internal phase transitions in monolayers
and bilayers, the logical next step would be to consider
internal phase transitions in lamellar stacks. They exist, 
of course, but they are presumably not very different from 
those in single bilayers. Thus we will shift focus and
concentrate on other aspects of membranes in this section:
shape fluctuations and defects~\cite{claire_diss,claire1,claire2}.

We consider a binary mixture of amphiphiles and solvent beads. 
The model shall not be described in full detail here. It was
originally introduced by Soddemann {\em et al}~\cite{toso,guo}
and has some similarity with the Smit model. The elementary
units are spheres with a hard core radius $\sigma$, which
may have two types: ``hydrophilic'' or ``hydrophobic''.
Beads of the same type attract each other at distances less 
than $1.5 \sigma$. ``Amphiphiles'' are tetramers made of 
two hydrophilic and two hydrophobic beads, and ``solvent'' 
particles are single hydrophilic beads. 

\begin{figure}[t]
\begin{center}
\epsfxsize=5cm
\epsfbox{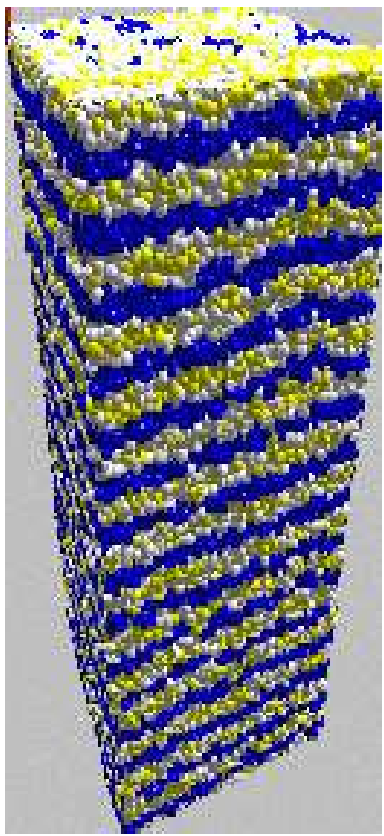}
\caption{\label{fig:claire_snapshot}
Snapshot of a bilayer stack with 30720 amphiphiles 
and 30720 solvent beads. The ``hydrophobic'' tail 
beads are blue, the ``hydrophilic'' head beads are 
white, and the solvent beads are yellow.
}
\end{center}
\end{figure}

The pressure and the temperature are chosen such that
the system is in a fluid lamellar phase. More specifically,
we study configurations with 5 or 15 stacked bilayers, which
are swollen with 20 volume percent solvent. 
The simulations were done with constant pressure molecular 
dynamics, using a Langevin thermostat to maintain constant 
temperature. The constant pressure algorithm was designed
such that the box dimensions parallel and perpendicular
to the lamellae fluctuate independently, in order to ensure 
that the pressure is isotropic and the membranes
have no surface tension. A configuration snapshot is shown 
in Fig.~\ref{fig:claire_snapshot}. It was produced by equilibration
of an initial configuration with lamellar order, set up such 
that the lamellae were oriented perpendicular to the long axis 
(the $z$-axis) of  the simulation box. We have checked 
with smaller systems that the lamellae self-assemble spontaneously 
if the initial configuration is disordered.

\begin{figure}[t]
\begin{center}
\epsfxsize=7.5cm
\epsfbox{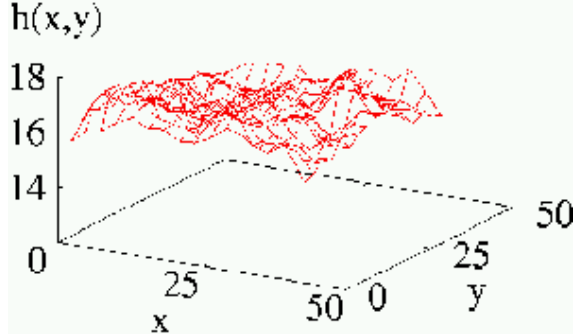}
\caption{\label{fig:membrane_snapshot}
Typical conformation of a membrane position $h_n(x,y)$.
}
\end{center}
\end{figure}

Our study aimed at analyzing shape fluctuations of the
membranes and defects in the membranes. Therefore, the
first nontrivial task was to determine the local positions 
of the membranes in the lamellar stack. This was done
as follows:
\begin{itemize}
\item 
  The simulation box was divided into $N_x N_y N_z$ cells.
  ($N_x = N_y = 32$). Note that the size of the cells
  may vary between configurations because the dimensions
  of the box fluctuate in a constant pressure simulation.
\item 
  In each cell, the relative density of tail beads 
  $\rho_{\mbox{\tiny tail}}(x,y,z)$ was calculated. It
  is defined as the ratio of the number of tail beads 
  and the total number of beads.
\item
  The hydrophobic space is defined as the space where
  the relative density of tail beads is higher than
  a given threshold $\rho_0$. The value of the threshold 
  is roughly 0.7 (80 \% of the maximum value of 
  $\rho_{\mbox{\tiny tail}}$ and depends on the mesh
  size.
\item 
  The cells that belong to the hydrophobic space are 
  connected to clusters. Two hydrophobic cells that share 
  at least one vortex are attributed to the same cluster. 
  Each cluster defines a membrane. This algorithm
  identifies membranes even if they have pores.
  At the presence of other membrane defects, 
  additional steps have to be taken. (This happened very
  rarely in our system).
\item 
  For each membrane $n$ and each position $(x,y)$,
  the two heights $h_n^{\mbox{\tiny min}}(x,y)$ 
  and $h_n^{\mbox{\tiny max}}(x,y)$, where the density
  $\rho_{\mbox{\tiny tail}}(x,y,z)$ passes through
  the threshold $\rho_0$, are determined. The mean 
  position is defined as the average 
$
   h_n(x,y) = 
    (h_n^{\mbox{\tiny min}} + h_n^{\mbox{\tiny max}})/2.
$
\end{itemize}
The algorithm identifies membranes even if they have pores. 
A typical membrane conformation $h_n(x,y)$ is shown in 
Fig.~\ref{fig:membrane_snapshot}.
The statistical distribution of $h_n(x,y)$ can
be analyzed and compared with theoretical predictions.

\begin{figure}[t]
\begin{center}
\epsfxsize=7cm
\epsfbox{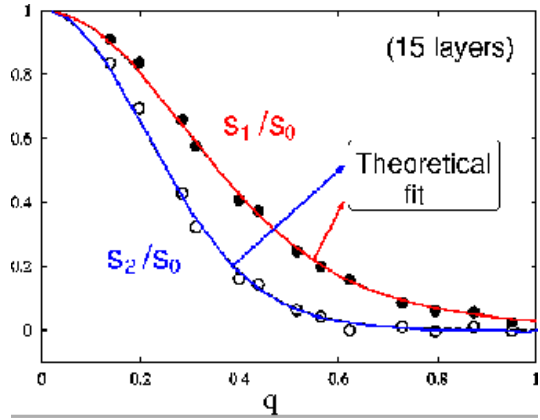}
\caption{\label{fig:trans_struc}
Ratios of transmembrane structure factors $s_1/s_0$ and $s_2/s_0$  
vs. in-plane wavevector $q$ in units of $\sigma^{-1}$.
The solid lines correspond to a theoretical fit to Eq.~(\ref{sn})
with one (common) fit parameter $K_c/B$.
}
\end{center}
\end{figure}

One of the simplest mesoscopic theories for fluctuations 
in membrane stacks is the ``discrete harmonic 
model''~\cite{lei}. 
It describes stacks of membranes without surface tension 
and assumes that the fluctuations are governed by two factors: 
The bending stiffness $K_c$ of single membranes, and the 
penalty for compressing the membranes, characterized by a 
compressibility modulus $B$. The free energy is given in 
harmonic approximation
\begin{equation}
\label{dh}
{\cal F}_{\mbox{\tiny DH}} = 
\sum_n \int_A dx \: dy \:
\Big\{ \frac{K_c}{2} 
(\frac{\partial^2 h_n}{\partial x^2} 
+ \frac{\partial^2 h_n}{\partial y^2})^2
+ \frac{B}{2} (h_n - h_{n+1} + \bar{d})^2
\Big\},
\end{equation}
where $\bar{d}$ is the average distance between layers.
This free energy functional is simple enough that statistical
averages can be calculated exactly. For example, the
transmembrane structure factor which describes correlations
between membrane positions in different membranes is
given by~\cite{claire1}
\begin{equation}
\label{sn}
s_n(q) = \langle h_m({\bf q})^* h_{m+n}({\bf q}) \rangle
= s_0(q) \: \Big[
1 + \frac{X}{2} - \frac{1}{2} \sqrt{X (X + 4)} \Big]^n,
\end{equation}
where $h_n({\bf q})$ is the Fourier transform of $h_n(x,y)$
in the $(x,y)$-plane and $X = q^4 K_c/B$ is a dimensionless parameter.
The function $s_0(q)$ can also be given explicitly~\cite{claire1}. 
We can test the prediction (\ref{sn}) by simply plotting the ratio
$s_n/s_0$ vs. $q$ for different $n$. The functional form of
the curves should be given by the expression in the square
brackets, with only one fit parameter $K_c/B$. 
Fig.~\ref{fig:trans_struc} shows our simulation data.
The agreement with the theory is very good over the whole range of 
$q$.  Hence the discrete harmonic model, a mesoscopic theory, 
describes the membrane fluctuations in our molecular way in a
satisfactory way.

Moreover, our analysis provides a value for the phenomenological
parameter $K_c/B$. By combining it with the analyses of other 
quantities, it is possible to determine $K_c$ and $B$ separately.
This establishes a bridge between the molecular and the mesoscopic 
description. 

Next we turn to discuss the membrane defects in the lamellar stack.
On principle, one can have three types of topological point defects 
in membranes: necks, pores, and passages. In our system, necks
and passages were extremely rare, and we did not collect enough
data to be able to analyze them statistically. Thus we will focus 
on the pores here. 

As we have already mentioned, the properties of pores determine
the permeability of a membrane. A number of atomistic
and coarse grained simulation studies have therefore addressed pore 
formation~\cite{marcus1,marcus2,marrink,zahn},
mostly in membranes under tension. In contrast, the membranes 
in our lamellar stack have no surface tension. As we shall see, 
this affects the characteristics of the pores quite dramatically.

\begin{figure}[t]
\begin{center}
\epsfxsize=6cm
\epsfysize=5.5cm
\epsfbox{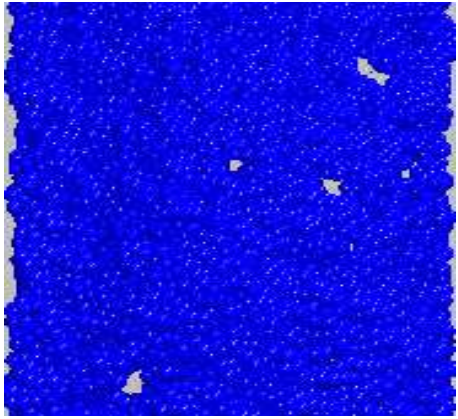}
\caption{\label{fig:pore_snapshot}
Snapshot of the hydrophobic beads in a single bilayer (top view).
}
\end{center}
\end{figure}

Fig.~\ref{fig:pore_snapshot} shows a snapshot of
a hydrophobic layer which contains a number of pores. These
pores have nucleated spontaneously. They ``live'' for a while,
grow and shrink without diffusing too much, until they finally
disappear. Most pores close very quickly, but some large ones 
stay open for a long time. 

\begin{figure}[b]
\begin{center}
\epsfxsize=6cm
\epsfbox{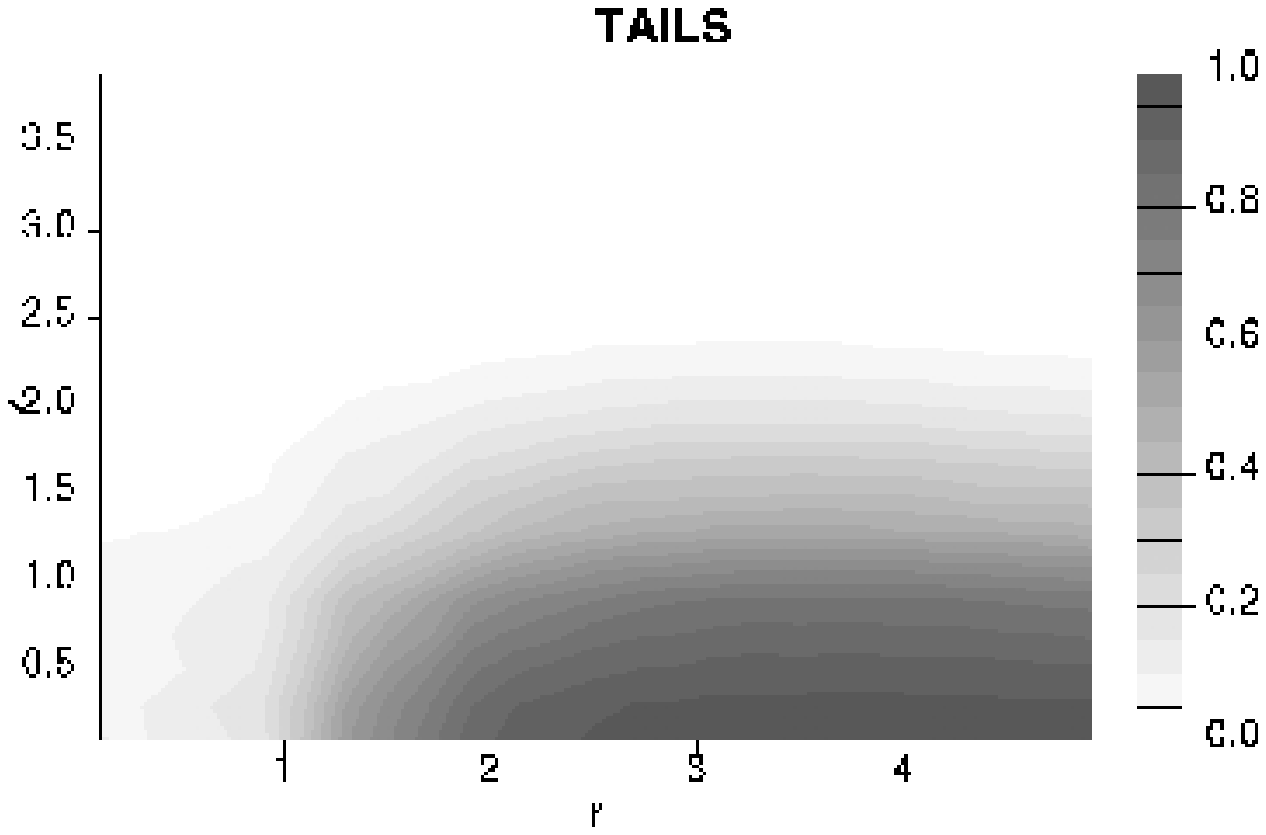}
\hfill
\epsfxsize=6cm
\epsfbox{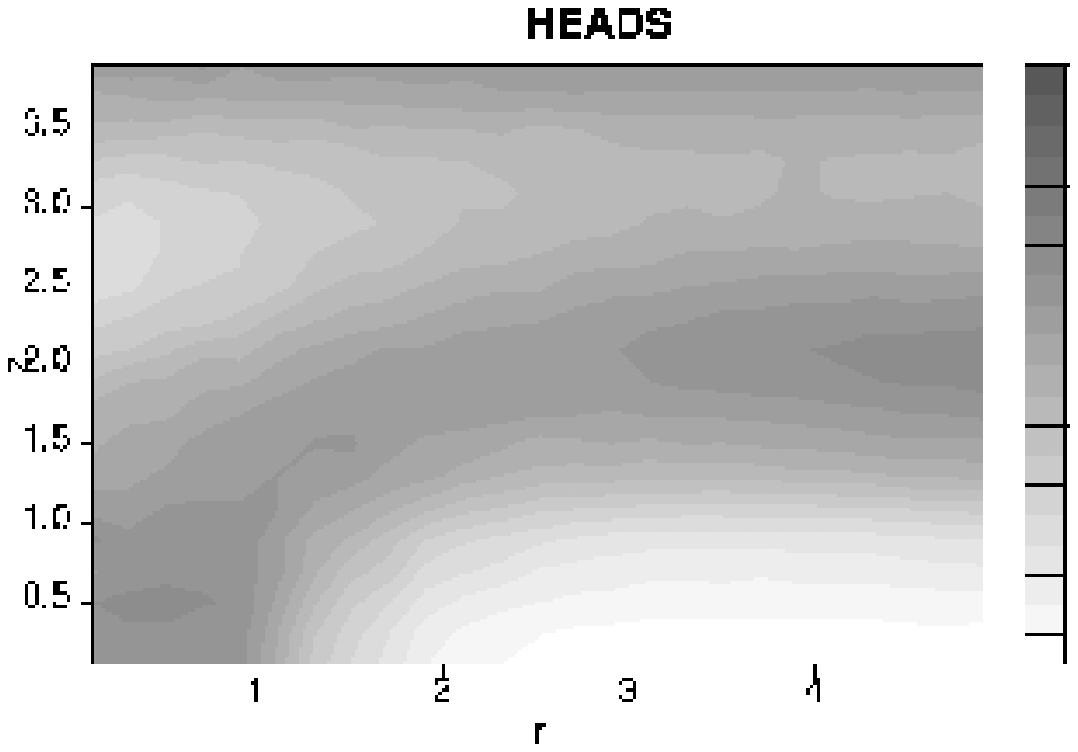}
\epsfxsize=6cm
\epsfbox{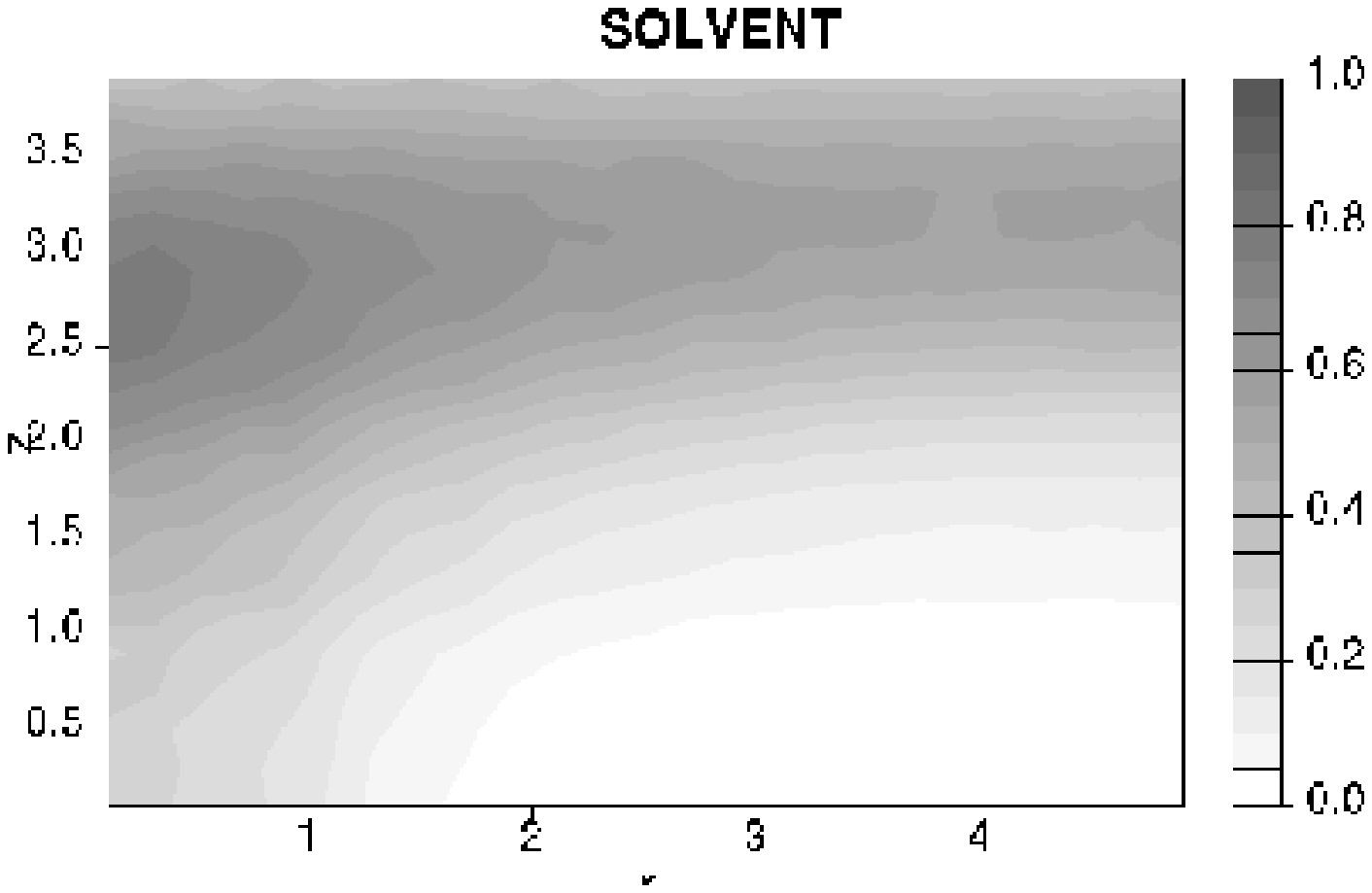}
\caption{\label{fig:density_profiles}
Density profiles of tail beads, head beads, and solvent
beads as a function of the distance from the center of the pore
in radial in-plane direction $r$ and normal direction $z$.
}
\end{center}
\end{figure}

We will first analyze the local membrane structure in the
vicinity of a pore. Fig.~\ref{fig:density_profiles} shows
average composition profiles of tail beads, head beads and
solvent beads as a function of the in-plane distance from 
the center of the pore, $r$, and the normal out-of-plane 
coordinate, $z$. These averages were performed with
data from pores with surface areas between 4 and 16 $\sigma^2$.
The pictures demonstrate that the amphiphiles rearrange
themselves at the pore edge, so that solvent beads
in the pore center are exposed mainly to head beads.
Such a pore is called hydrophilic. In previous simulations,
both hydrophilic and hydrophobic pores have been reported,
depending on the system under consideration.

\begin{figure}[t]
\begin{center}
\epsfxsize=7cm
\epsfbox{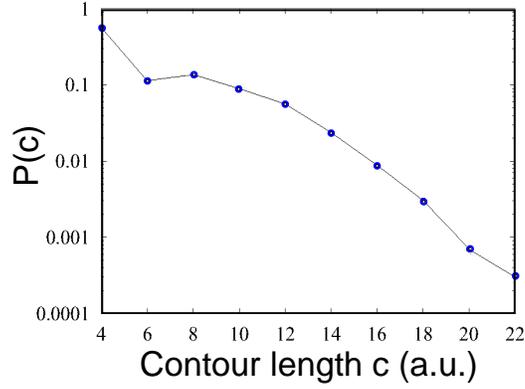}
\caption{\label{fig:hist_contours1}
Distribution of pore contours in a semi-logarithmic plot.
}
\end{center}
\end{figure}

Whereas the local composition profiles at the pore edge depend 
on the model, other structural properties of pores on
larger scales are presumably generic and can again be 
described by simple mesoscopic theories. For example, the 
simplest Ansatz for the free energy of a pore with the area $A$ 
and the contour length $c$ has the form~\cite{lister}
\begin{equation}
\label{pore_energy}
E = E_0 + \lambda \: c - \gamma \: A,
\end{equation}
where $E_0$ is a core energy, $\lambda$ a material parameter
called line tension, and $\gamma$ the surface tension.
The last term accounts for the reduction of energy due
to the release of surface tension in a stretched membrane.
In our case, the surface tension is zero ($\gamma = 0$), and
the last term vanishes.
The second term describes the energy penalty at the pore rim.
If this simple free energy model is correct, the pore 
shapes should be distributed according to a Boltzmann
distribution 
\begin{equation}
\label{pc_1}
P(c) \propto \exp(- \lambda c).
\end{equation}
This can be tested in the simulation. 
Fig.~\ref{fig:hist_contours1} shows a histogram of contour 
lengths $P(c)$. The bare data do not reflect the expected 
exponential behavior. Either the Ansatz (\ref{pore_energy}) is
wrong, or we have not used it correctly.

Indeed, a closer look reveals that Eq.~(\ref{pc_1}) disregards
an important effect: The ``free energy'' (\ref{pore_energy}) 
gives only {\em local} free energy contributions, 
stemming from local interactions and local amphiphile rearrangements. 
In addition, one must also account for the {\em global} entropy of 
possible contour length conformations. Therefore, we have to evaluate 
the ``degeneracy'' of contour lengths $g(c)$, and test the 
relation
\begin{equation}
\label{pc_2}
P(c) \propto g(c) \exp(- \lambda c).
\end{equation}

\begin{figure}[t]
\begin{center}
\epsfxsize=7cm
\epsfbox{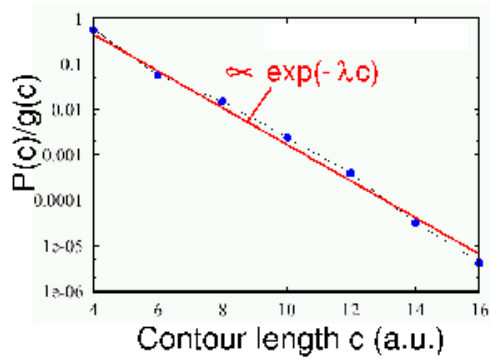}
\caption{\label{fig:hist_contours2}
Distribution of pore contours, divided by the degeneracy function $g(c)$.
}
\end{center}
\end{figure}

Fig.~\ref{fig:hist_contours2} demonstrates that this second 
Ansatz describes the data very well. From the linear fit to the 
data, one can extract a value for the line tension $\lambda$. 
Thus we have again established a bridge between the molecular
simulation model and a mesoscopic theory.

\begin{figure}[b]
\begin{center}
\epsfxsize=7cm
\epsfbox{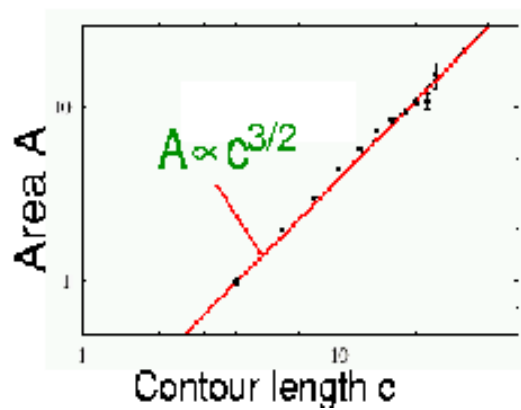}
\caption{\label{fig:area_contour}
Pore area vs. contour length (arbitrary units).
}
\end{center}
\end{figure}

If the model (\ref{pore_energy}) is correct, it makes a 
second important prediction: Since the free energy only depends
on the contour length, pores with the same contour length are
equivalent and the shapes of these pores should be distributed
like those of two dimensional ring polymers.
In particular, they are not round, but have a fractal structure.
From polymer theory, one knows that the size $R_g$ of a two 
dimensional self-avoiding polymer scales roughly like 
$R_g \propto N^{3/4}$ with the polymer length $N$. In our case,
the ``polymer length'' is the contour length $c$. 
Thus the area $A$ of a pore should scale like
\begin{equation}
A \propto R_g^2 \propto ( C^{3/4}) ^2 = C^{3/2}.
\end{equation}
Fig.~\ref{fig:area_contour} shows that this is indeed the case.

Many other properties of the pores can be investigated, {\em e.~g.},  
pore distributions, the dynamical evolution of pores, pore life 
times~\cite{claire_diss,claire2}. 
Nevertheless, we shall stop here. We hope that our brief discussion
has conveyed an idea how a coarse-grained molecular simulation 
help to test and justify mesoscopic theories, and to establish
a connection between molecular and mesoscopic descriptions of 
amphiphilic systems.

\section{Outlook}
\label{sec:outlook}

Simulations like those presented here are only a very first step 
towards understanding the properties of membranes. First, real 
biomembranes are not pure systems, but contain a mixture of many 
different lipids, with saturated or unsaturated chains, with charged 
or neutral heads etc.  Second, biomembranes are filled with proteins. 
Typical biomembranes are not homogeneous, but compartmented into 
several regions with different lipid and protein composition. 
Furthermore, biomembranes have a complex environment, which 
influences the membrane properties. Finally, biomembranes are not
equilibrium structures. They contain active proteins, and they are
surrounded by an ever changing environment. 

These many complications seem discouraging. We hope to have 
shown that a systematic approach, where the different aspects of 
membranes are studied one by one with appropriated idealized models, 
can be rewarding. Still, much remains to be done. Computer simulations 
of membranes and biomembranes will certainly be an active and lively 
research area for a long time.

\section*{Acknowledgments}

We have enjoyed collaborations with C.~Stadler, H.~Lange, 
M.~Mareschal, and K.~Kremer.  We have benefitted from
fruitful discussions with F.~M.~ Haas, R.~Hilfer, K.~Binder,
T.~Soddemann, H.-X.~Guo, and R.~Everaers. The simulations were 
carried out at the NIC J\"ulich, the computing center of the
Max-Planck society in Garching, and the computing center of the
Commissariat \`a l'Energie Atomique (Grenoble). We acknowledge
financial support by the German Science Foundation and
by the ``R\'egion Rh\^one-Alpes''.

\end{document}